\documentclass[11pt]{article}
\usepackage{amsmath}
\usepackage{amsfonts,amssymb,latexsym}

\begin{document}

\newcommand{\nl}{\nonumber\\}
\newcommand{\nnl}{\nl[6mm]}
\newcommand{\nle}{\nl[-2.5mm]\\[-2.5mm]}
\newcommand{\nlb}[1]{\nl[-2.0mm]\label{#1}\\[-2.0mm]}

\renewcommand{\leq}{\leqslant}
\renewcommand{\geq}{\geqslant}

\newcommand{\be}{\bes}
\newcommand{\ee}{\ees}
\newcommand{\bes}{\begin{eqnarray}}
\newcommand{\ees}{\end{eqnarray}}
\newcommand{\eens}{\nonumber\end{eqnarray}}

\renewcommand{\/}{\over}
\renewcommand{\d}{\partial}
\newcommand{\ddx}{{d\/dx}}

\newcommand{\vect}{{\mathfrak{vect}}}

\newcommand{\mn}{{\mu\nu}}

\newcommand{\e}{{\mathrm e}}
\newcommand{\emx}{\e^{im\cdot x}}

\newcommand{\la}{\lambda}

\newcommand{\Lmu}{L_\mu}
\newcommand{\Lnu}{L_\nu}

\newcommand{\ZZ}{{\mathbb Z}}

\title{Virasoro 3-algebra from scalar densities}

\author{T. A. Larsson \\
Vanadisv\"agen 29, S-113 23 Stockholm, Sweden\\
email: thomas.larsson@hdd.se}

\maketitle
\begin{abstract}
It is shown that the ternary Virasoro-Witt algebra of Curtright, 
Fairlie and Zachos can
be constructed by applying the Nambu commutator to the $\vect(1)$ 
realization on scalar densities. This construction is generalized to
$\vect(d)$, but the corresponding 3-algebra fails to close.
\end{abstract}

\bigskip

There has recently been a surge of interest in 3-algebras in M-theory
\cite{BL,GMR,Gus,HHM}, which is closely related to the ternary
brackets introduced long ago by Nambu \cite{Nam} and developed by
Filippov \cite{Fil}, Curtright and Zachos \cite{CZ02} and many others.
In particular, Lin \cite{Lin} considered a Kac-Moody 3-algebra and
Curtright, Fairlie and Zachos \cite{CFZ08} considered
very recently a 3-algebra related to the Witt (centerless Virasoro)
algebra. In this note I observe that their Virasoro-Witt 3-algebra 
(eqn (22) in \cite{CFZ08}) can be constructed by applying the Nambu 
commutator
\bes
[A,B,C] &=& ABC + BCA + CAB - BAC - CBA - ACB 
\nlb{Nambu}
&=& A[B,C] + B[C,A] + C[A,B]
\eens
to the Virasoro representation acting on scalar densities, i.e. primary 
fields.

Consider the operators
\bes
E_m &=& \e^{imx}, \nl
L_m &=& \e^{imx}(-i\ddx + \la m), 
\label{def}\\
S_m &=& \e^{imx}(-i\ddx + \la m)^2.
\eens
They satisfy
\bes
E_m E_n &=& E_{m+n}, \nl
E_m L_n &=& L_{m+n} - \la m E_{m+n}, 
\nlb{prod}
L_m E_n &=& L_{m+n} + (1-\la) n E_{m+n}, \nl
L_m L_n &=& S_{m+n} + (n - \la(m+n))L_{m+n} + (\la^2 - \la)mnE_{m+n}.
\eens
The products involving $S$'s can also be readily computed but are not
needed here. The commutators $[A,B] = AB - BA$ are
\bes
[E_m, E_n] &=& 0, \nl
{[}L_m, E_n] &=& n E_{m+n}, 
\label{Vir}\\
{[}L_m, L_n] &=& (n-m)L_{m+n}.
\eens
This is recognized as the semidirect product of the Witt algebra
$\vect(1)$ (centerless Virasoro algebra) with an abelian current algebra.
The corresponding Nambu commutators (\ref{Nambu}) become
\bes
[E_m,E_n,E_r] &=& 0, \nl
{[}L_m,E_n,E_r] &=& (n-r)E_{m+n+r}, 
\nlb{Witt3}
{[}L_m,L_n,E_r] &=& (n-m)L_{m+n+r} 
 + (1-2\la)(n-m)rE_{m+n+r}, \nl
{[}L_m,L_n,L_r] &=& (\la-\la^2)(m-n)(n-r)(r-m)E_{m+n+r}.
\eens
This is the Virasoro-Witt 3-algebra \cite{CFZ08}, eqn (22). To recover
their conventions, we substitute $L_m \to -L_m$, $E_m \to M_m$, 
$\la \to \beta$, and note that their constant $C = \beta(1-\beta)$, cf
their eqn (10). We have thus shown that the Virasoro-Witt 3-algebra
reported by Curtright, Fairlie and Zachos can be obtained by inserting
the $\vect(1)$ representation on scalar densities into the Nambu
commutator.

It was noted by those authors that the 3-algebra (\ref{Witt3}) does not
satisfy the fundamental identity of type $LLLLE$, except when 
$\la = \pm 2i$. However, it was emphasized earlier \cite{CZ02} that
there is nothing fundamental about the fundamental identity as far as
associativity is concerned. Indeed, since we started from an
infinite-dimensional matrix represention (\ref{def}) of $\vect(1)$, the
3-algebra (\ref{Witt3}) does possess an associative matrix
representation by construction.

Let us generalize this construction to higher dimensions; instead of the
algebra $\vect(1)$ of vector fields on the circle, consider the algebra
$\vect(d)$ of vector fields on the $d$-dimensional torus. Let 
$x = (x^\mu)$ be the coordinates and $\d_\mu = \d/\d x^\mu$ be the
corresponding derivatives. The Fourier modes $\emx$ are labelled by
momenta $m = (m_\mu) \in \ZZ^d$. For simplicity, we restrict ourselves
to scalar fields, i.e. densities with weight $\la=0$. Consider in
analogy with (\ref{def})
\bes
E(m) &=& \emx, \nl
\Lmu(m) &=& \emx(-i\d_\mu), 
\label{ddef}\\
S_\mn(m) &=& S_{\nu\mu}(m)\ =\ \emx (-i\d_\mu)(-i\d_\nu).
\eens
These operators satisfy the products
\bes
E(m)E(n) &=& E(m+n), \nl
E(m)\Lnu(n) &=& \Lnu(m+n), \nle
\Lmu(m)E(n) &=& \Lmu(m+n) + n_\mu E(m+n), \nl
\Lmu(m)\Lnu(n) &=& S_\mn(m+n) + n_\mu\Lnu(m+n).
\eens
The commutators
\bes
[E(m),E(n)] &=& 0, \nl
{[}\Lmu(m), E(n)] &=& n_\mu E(m+n), \\
{[}\Lmu(m), \Lnu(n)] &=& n_\mu \Lnu(m+n) - m_\nu\Lmu(m+n),
\eens
satisfy a semidirect product of $\vect(d)$ with an abelian
current algebra in $d$ dimensions.
The Nambu commutators become
\bes
[E(m),E(n),E(r)] &=& 0, \nl
{[}\Lmu(m),E(n),E(r)] &=& (n_\mu - r_\mu) E, 
\label{3vectd}\\
{[}\Lmu(m),\Lnu(n),E(r)] &=& 
(m_\nu + r_\nu)\Lmu - (n_\mu + r_\mu)\Lnu 
+ (n_\mu r_\nu - m_\nu r_\mu) E, \nl
{[}\Lmu(m),\Lnu(n),L_\rho(r)] &=&
(m_\rho-n_\rho)S_\mn + (n_\mu-r_\mu)S_{\nu\rho} 
+ (r_\nu-m_\nu)S_{\rho\mu} \nl
\qquad+ (n_\mu r_\nu - m_\nu r_\mu)L_\rho
&+& (r_\nu m_\rho - n_\rho m_\nu)\Lmu
+ (m_\rho n_\mu - r_\mu n_\rho)\Lnu,
\eens
where the common argument $(m+n+r)$ on the RHS has been suppressed.
Unlike the Witt 3-algebra (\ref{Witt3}), these brackets do not close,
due to the appearence of $S$ terms in the $LLL$ bracket. Hence we must
also consider brackets with $S$'s in the LHS, which leads to infinite
hierarchy of new generators.

In the one-dimensional case we started from the realization (\ref{def})
on scalar densities, but in $d$ dimensions we only considered the case
$\la=0$. The $\vect(d)$ realization (\ref{ddef}) can readily be 
generalized to scalar densities:
\be
\Lmu(m) &=& \emx (-i\d_\mu + \la m_\mu).
\label{scalar}
\ee
More generally, the classical irreps of $\vect(d)$ are closely related
to tensor densities, which correspond to the $\vect(d)$ realization
\be
\Lmu(m) &=& \emx (-i\d_\mu + m_\nu T^\nu_\mu),
\label{tensor}
\ee
where $T^\mu_\nu$ are some matrices which satisfy $gl(d)$:
\be
[T^\mu_\nu, T^\rho_\sigma] &=& \delta^\rho_\nu T^\mu_\sigma
- \delta^\mu_\sigma T^\rho_\nu.
\ee
For each $gl(d)$ representation $R$, these formulas yield a $\vect(d)$ 
representation, which describes how infinitesimal diffeomorphisms act
on tensor densitites of type $R$. In particular, we recover scalar
densitites by setting
\be
T^\mu_\nu = \la \delta^\mu_\nu. 
\ee
The tensor representation (\ref{tensor}) is irreducible, except when
$R$ is totally antisymmetric with weight zero; in that case, the module
of differential forms has the submodule of closed forms. There are
essentially no other irreducible $\vect(d)$ representations
\cite{Rud74}.

We can now construct new $\vect(d)$ 3-algebras by replacing the
definition of $\Lmu(m)$ in (\ref{ddef}) by (\ref{scalar}) or
(\ref{tensor}), and change the definition of $S_\mn(m)$ accordingly.
However, since the $\vect(d)$ 3-algebra (\ref{3vectd}) fails to close
already when we start from scalar fields, we must introduce an infinite
hierarchy of new generators and relations. The situation becomes
very complex, and the physical relevance of such an exercise is unclear
to me.

\end{document}